\documentclass[a4paper,fleqn]{cas-dc}
\usepackage[numbers]{natbib}
\usepackage{threeparttablex, booktabs, url, amsmath, xr, nature_abbr, flushend} %

\def\tsc#1{\csdef{#1}{\textsc{\lowercase{#1}}\xspace}}
\tsc{WGM}
\tsc{QE}
\def\pin{$\mathcal{PI}$ }
\def\mle{$\mathcal{MLE}$ }
\def\mae{$\mathcal{MAE}$ }

\begin{document}

\shorttitle{Zhang et al.}

\title [mode = title]{The arrival time and energy of FRBs traverse the time-energy bivariate space like a Brownian motion}

\author[1,2,3]{Yong-Kun Zhang} %

\author[1,2,3,4,5]{Di Li} %
\cormark[1]
\ead{dili@nao.cas.cn}

\author[4]{Yi Feng} %
\author[1,2,6]{Pei Wang} %
\author[7]{Chen-Hui Niu} %
\author[8]{Shi Dai} %
\author[9,10,11]{Ju-Mei Yao} %
\author[1,2,3,6]{Chao-Wei Tsai} %

\affiliation[1]{organization={National Astronomical Observatories},
            addressline={Chinese Academy of Sciences}, 
            city={Beijing},
            citysep={}, %
            postcode={100101}, 
            country={China}}

\affiliation[2]{organization={Key Laboratory of Radio Astronomy and Technolgoy},
            addressline={Chinese Academy of Sciences}, 
            city={Beijing},
            citysep={}, %
            postcode={100101}, 
            country={China}}

\affiliation[3]{organization={University of Chinese Academy of Sciences},
            city={Beijing},
            citysep={}, %
            postcode={100049}, 
            country={China}}

\affiliation[4]{organization={Research Center for Astronomical Computing},
            addressline={Zhejiang Laboratory}, 
            city={Hangzhou},
            citysep={}, %
            postcode={311100}, 
            country={China}}

\affiliation[5]{organization={New Cornerstone Science Laboratory},
            city={Shenzhen},
            citysep={}, %
            postcode={518054}, 
            country={China}}

\affiliation[6]{organization={Institute for Frontiers in Astronomy and Astrophysics},
            addressline={Beijing Normal University}, 
            city={Beijing},
            citysep={}, %
            postcode={102206}, 
            country={China}}

\affiliation[7]{organization={Institute of Astrophysics},
            addressline={Central China Normal University}, 
            city={Wuhan},
            citysep={}, %
            postcode={430079}, 
            country={China}}

\affiliation[8]{organization={School of Science, Western Sydney University},
            addressline={Locked Bag 1797}, 
            city={Penrith NSW},
            citysep={}, %
            postcode={2751}, 
            country={Australia}}

\affiliation[9]{organization={Xinjiang Astronomical Observatory},
            addressline={Chinese Academy of Sciences}, 
            city={Urumqi},
            citysep={}, %
            postcode={830011}, 
            country={China}}

\affiliation[10]{organization={Key Laboratory of Radio Astronomy},
            addressline={Chinese Academy of Sciences}, 
            city={Urumqi},
            citysep={}, %
            postcode={830011}, 
            country={China}}

\affiliation[11]{organization={Xinjiang Key Laboratory of Radio Astrophysics},
            city={Urumqi},
            citysep={}, %
            postcode={830011}, 
            country={China}}

\cortext[1]{Corresponding author}

\begin{abstract}
The origin of fast radio bursts (FRBs), the brightest cosmic explosion in radio bands, remains unknown. We introduce here a novel method for a comprehensive analysis of active FRBs' behaviors in the time-energy domain. Using ``Pincus Index'' and ``Maximum Lyapunov Exponent'', we were able to quantify the randomness and chaoticity, respectively, of the bursting events and put FRBs in the context of common transient physical phenomena, such as pulsar, earthquakes, and solar flares. In the bivariate time-energy domain, repeated FRB bursts' behaviors deviate significantly (more random, less chaotic) from pulsars, earthquakes, and solar flares. The waiting times between FRB bursts and the corresponding energy changes exhibit no correlation and remain unpredictable, suggesting that the emission of FRBs does not exhibit the time and energy clustering observed in seismic events. The pronounced stochasticity may arise from a singular source with high entropy or the combination of diverse emission mechanisms/sites. Consequently, our methodology serves as a pragmatic tool for illustrating the congruities and distinctions among diverse physical processes.
\end{abstract}

\begin{keywords}
Fast Radio Burst \sep  Earthquake \sep Stochastic \sep Chaos
\end{keywords}

\maketitle

\section{Introduction}

Fast radio bursts (FRBs) are intense pulses of radio emission that last just a few milliseconds. First discovered in 2007 \cite{2007Sci...318..777L}, FRBs have since been observed by a variety of radio telescopes around the world. Despite their ubiquity, however, the origin of these mysterious signals remains unknown. FRBs have a wealth of observational parameters carrying information about the sources and the propagation paths, including arrival time, energy, duration, bandwidth, polarization, dispersion, scintillation, scattering, etc. Recent studies on the polarization of FRBs suggest that FRBs are located in complex magnetized environments \cite{2022MNRAS.510.4654B, 2022Sci...375.1266F, 2022Natur.609..685X, 2023Sci...380..599A}. These studies demonstrate information about the propagation paths of FRBs. Time and energy are two observational parameters directly related to the radiation nature of FRBs. The observation of FRB~20121102A with the Five-hundred-meter Aperture Spherical radio Telescope (FAST) for the first time revealed a bimodal distribution of FRB energy, suggesting that there may be different radiation mechanisms for FRBs \cite{li21, 2021ApJ...919...89Y}.

The study of time and energy sequences of repeating FRB bursts enables a deeper understanding of the origin of FRBs, which is difficult to achieve with non-repeating bursts because their only burst is just a point in time-energy phase space. Here, we use two active repeating FRBs as our analysis objects, which are the only two FRBs known to have associated compact persistent radio sources (PRSs) \cite{chatterjee17, 2017ApJ...834L...8M, 2022Natur.606..873N}, namely FRB~20121102A and FRB~20190520B. Both have gone through highly active episodes over a substantial dynamic range of time scales (milliseconds to months), as detected by FAST \cite{li21, 2022Natur.606..873N}. Due to the high sensitivity and high cadence coverage of FAST's observations of these two FRBs, we can obtain more complete event series than other telescopes. We report here a first systematic examination of the FRB behaviors in the time-energy bivariate domain.

\section{Data}

\begin{figure*}[!htp]
    \centering
    \includegraphics[width=0.8\textwidth]{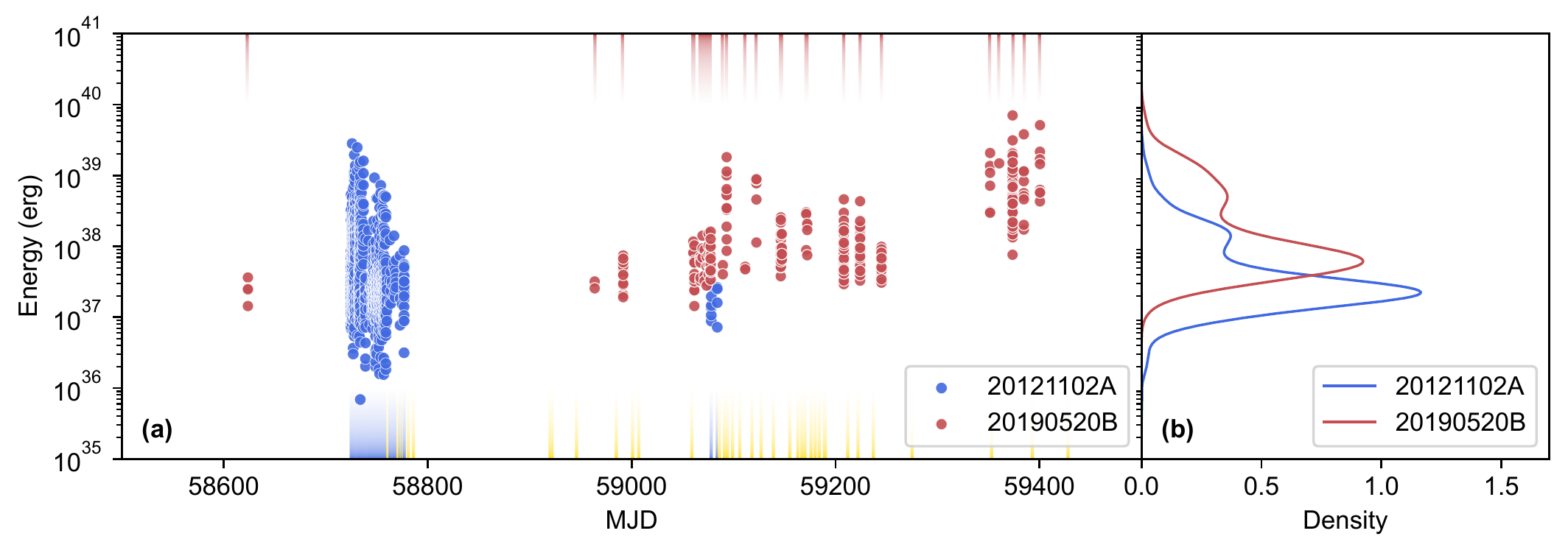}
    \caption{{\bf The detected bursts with energy from FRB~20121102A and FRB~20190520B.} In {\bf panel a}, the blue dots and red dots show the energy and arrival time of bursts from the two FRBs. The blue and red bars are the observation session of the two FRBs with detection, the yellow bars are the observation session without detection. {\bf Panel b} shows the kernel density estimate (KDE) energy distribution of the two FRBs.}
    \label{fig:ob}
\end{figure*}

FRB~20121102A was the first FRB found to be repeating and to be precisely localized \cite{spitler16, chatterjee17, tendulkar17}. In one extremely active epoch, FAST detected 1652 pulses from FRB~20121102A between August 29 and October 29, 2019 \cite{li21}. Since then, we have been carrying out regular monitoring of FRB~20121102A once every one or two months. On August 17 and 23, 2020, FAST caught another 12 bursts. Since then, no more burst has been detected. FRB~20190520B is the first repeating FRB discovered by FAST during the Commensal Radio Astronomy FAST Survey (CRAFTS) \cite{li19}, which has been accurately located with a PRS \cite{2022Natur.606..873N}. Since its discovery on May 20th, 2019, more than 200 pulses have been detected by FAST and Parkes \cite{2023Sci...380..599A}. Fig. \ref{fig:ob} shows the observation coverage and the detected bursts of FRB~20121102A and FRB~20190520B.

\section{Time domain analysis and results}

\begin{figure}[!htp]
    \centering
    \includegraphics[width=0.47\textwidth]{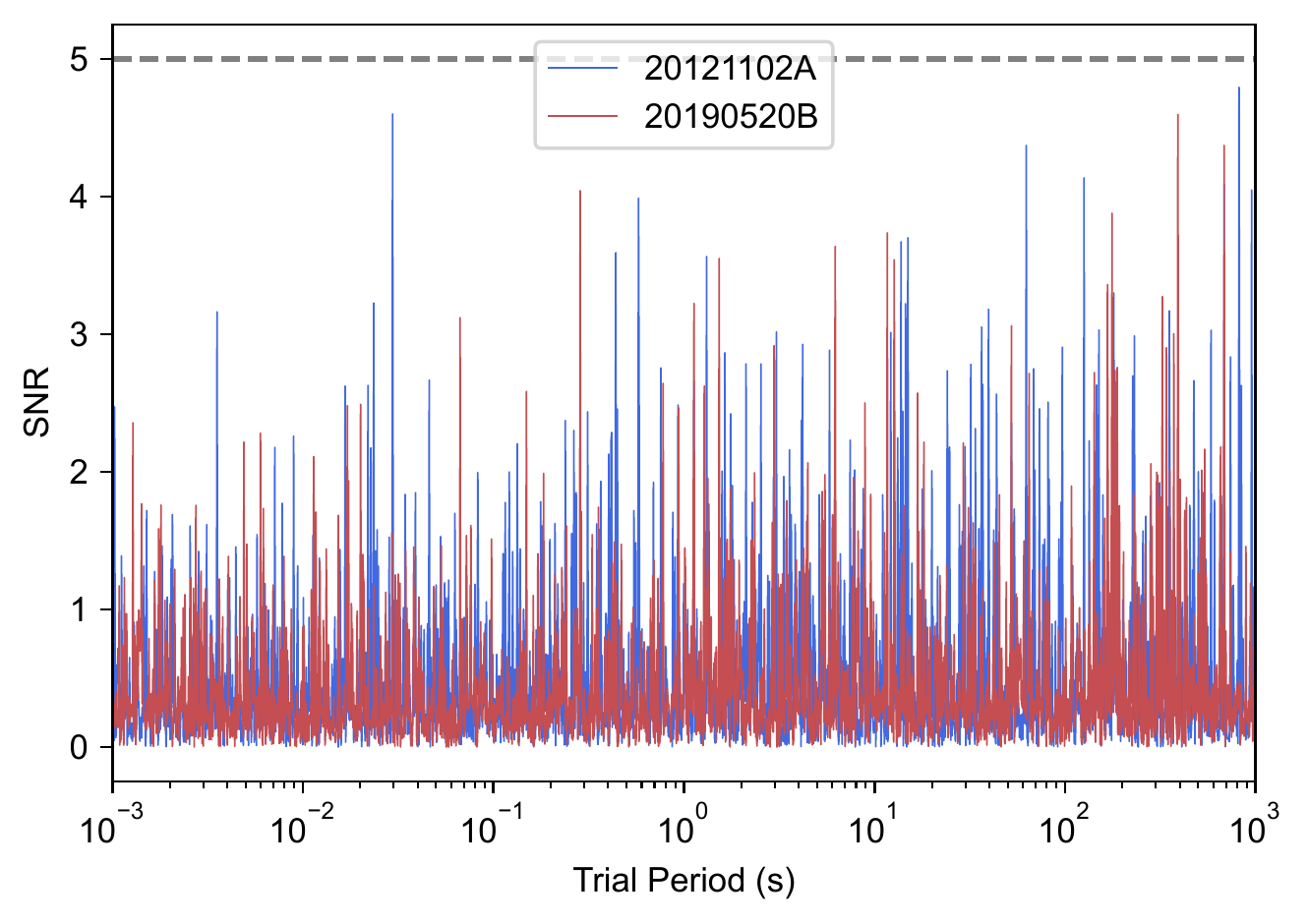}
    \caption{{\bf The periodogram of FRB~20121102A and FRB~20190520B.}}
    \label{fig:sp}
\end{figure}

We employed the Lomb-Scargle Periodogram (LSP) \cite{lomb, scargle} to search for periods of FRB~20121102A and FRB~20190520B, yet no statistically significant period signals were identified within the range of 1~ms to 1000~s. To quantify the significance (signal-to-noise ratio, S/N) of signals at different periods, we compared the power differences between randomly generated time series and the time series of FRB bursts in the LSP. For FRB~20121102A, we generated a time series with the same number as the FRB bursts between the time of the first and last burst's arrival time, drawn from a uniform distribution. Subsequently, we applied the LSP to this randomly generated sequence. We repeated this process 100 times, resulting in a distribution of periodogram powers for random signals at different periods. We conducted the same procedure for FRB~20190520B. Within the range of 1~ms to 1000~s, no significant periodic signals for both FRBs surpassed $5\sigma$ of the distribution of periodogram powers from the random time series (Fig.~\ref{fig:sp}). Thus we have significantly constrained the possibility of periodic signal existence. Several effects, including variable emission altitudes/sites, may complicate the inference of a magnetar's spin period from the arrival times of FRB bursts \cite{2022ApJ...929...97C}. An alternative interpretation could be that the period of the magnetar powering FRB is much longer than that of Galactic magnetars, so there is not short-timescale periodicity in the FRB data \cite{2020MNRAS.496.3390B}. Regardless of the specific cause, the absence of a detected periodic signal still highlights the inherent randomness in the emission of FRBs.

We now look into the more fundamental aspects of FRBs' time-domain behaviors through analysis of waiting times. The waiting time between two events is $\delta=t_{i+1}-t_i$, where $t_{i+1}$ and $t_{i}$ are the arrival times for the $(i+1)_{th}$ and $i_{th}$ events, respectively. 

Since the discovery of FRB~20200428, the origin of FRBs being from magnetars has gradually gained popularity \cite{2020Natur.587...54C, 2020Natur.587...59B}. As a possible source model for FRBs, the production of FRBs by magnetars relies on some trigger mechanisms \cite{2023RvMP...95c5005Z}, including crust cracking at the neutron star surface \cite{2019ApJ...879....4W, wang18, 2021ApJ...919...89Y}, sudden magnetic reconnection events in the magnetosphere \cite{2010vaoa.conf..129P}, or triggers from external events \cite{2020ApJ...897L..40D, 2017ApJ...836L..32Z}. Motivated by this, we compared FRBs with earthquake and solar flare with similar but not identical mechanisms.

The seismic data is from Southern California from the Southern California Earthquake Data Center \cite{earthquake10}, which contains information such as the occurrence time, latitude, longitude, and magnitude of the earthquake since 1932. For earthquake, we selected events within a region of  $2^\circ\times2^\circ$ as a continuous seismic sequence. The magnitudes of all earthquakes are converted into energy (erg) through an empirical relation \cite{bath66}. Solar flare data is from the Hinode Flare Catalogue \cite{hinode}. We simulated a 100-step Brownian motion. In mathematics, Brownian motion is described by the Wiener process, a stochastic process $W(t)$ concerning time $t$. According to the definition of Brownian motion, for time $t$ and $s$, the increments of Brownian motion $W(t)-W(s)$ follow a normal distribution $\mathcal{N}(0, t-s)$. Therefore, our simulation proceeded as follows: firstly, we sampled the waiting time $\delta t$ for each step of the Brownian motion from the exponential distribution $P(t) = \lambda e^{-\lambda t}$, where $\lambda=1$, corresponding to a Poisson process with a event rate of 1. Subsequently, based on each step's waiting time, we sampled the step size from the normal distribution $\mathcal{N}(0, dt)$.

\begin{figure}[!htp]
    \centering
    \includegraphics[width=0.47\textwidth]{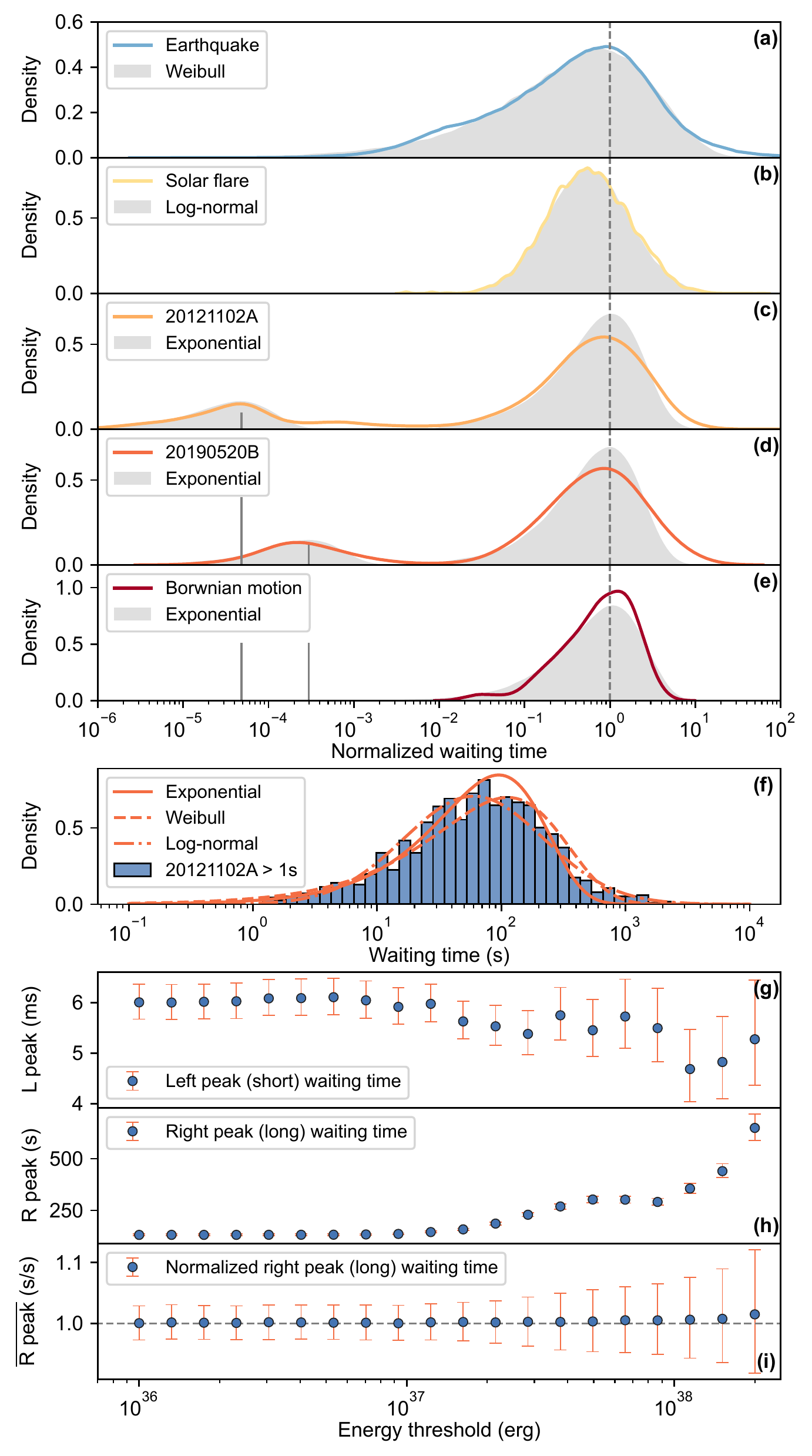}
    \caption{{\bf The normalized waiting time distribution of earthquake, solar flare, two FRBs, and Brownian motion.} Colored lines in {\bf panel a-e} are the KDEs of waiting times for these phenomena, which are normalized by $\bar T$ of 3435.1~s, 25810.4~s, 88.4~s, 123.3~s, and 190.4~s, respectively. Gray regions in panel a-e indicate the fitting of waiting times with Weibull function, LogNormal function, and Exponential function (Poisson process). The gray solid lines denote the short characteristic waiting time of the two FRBs, which are $\sim 4$~ms for FRB~20121102A and $\sim 36$~ms for FRB~20190520B. {\bf Panel f} shows the distribution of waiting times longer than 1 s for FRB 20121102A. The solid line, dashed line, and dash-dot line respectively denote the fitting results of the exponential function, Weibull function, and lognormal function. {\bf Panel g-i} show the left peak, right peak and normalized right peak of waiting time under different energy thresholds.}
    \label{fig:wt}
\end{figure}

The waiting time distributions of these phenomena are presented in Fig. \ref{fig:wt}a-e. We defined an average time as $\bar T = {\sum T_i}/{\sum N_i}$, where $T_i$ and $N_i$ are the length and event count of the observation session. All waiting times are normalized by corresponding $\bar T$. 
For fitting the waiting time distribution, we employed the \textsc{emcee} package to perform maximum likelihood estimation on the fitting parameters. We define the likelihood function as 
\begin{equation}
    L(\theta|t)=\sum_i\log f(t, \theta),
\end{equation}
where $t$ is the waiting times, $f(t, \theta)$ is the model for waiting time distribution, and $\theta$ is the free parameter(s) for the model. $f(t, \theta)$ could be
\begin{itemize}
    \item Weibull distribution \\
    $f(t, \theta) = f(t, k, \lambda)=\frac k\lambda\left(\frac t\lambda\right)^{k-1}e^{-(t/\lambda)^k}$ \\
    where $\theta$ represents the shape parameter $k$ and the scale parameter $\lambda$.
    \item Log-normal distribution \\
    $f(t, \theta) = f(t, \mu,\sigma) = \frac {1}{x\sigma\sqrt{2\pi}}\exp\left[-\frac{(\ln x-\mu)^2}{2\sigma^2}\right]$ \\
    where $\theta$ represents the expected value $\mu$ and standard deviation $\sigma$ of the variable's natural logarithm.
    \item Exponential distribution \\
    $f(t, \theta) = f(t, \lambda)=\lambda e^{-\lambda t}$ \\
    where $\theta$ represents the event rate $\lambda$.
    \item Mixture of two exponential distributions \\
    $f(t, \theta) = f(t, \lambda_1, \lambda_2, p)=p e^{-\lambda_1 t} + (1-p/\lambda_1) \lambda_2 e^{-\lambda_2 t}$ \\
    where $\theta$ representing two event rates $\lambda_{1,2}$ and a proportionality factor $p$.
\end{itemize}

The waiting time distribution for a Poisson process follows an exponential distribution. The waiting time distribution of the earthquake deviates from the exponential distribution and can be better described by a Weibull distribution. This means that earthquake events with shorter waiting times occur more frequently than expected from a simple Poisson process, i.e. clustering in time. The waiting time distribution of solar flare can be fitted by a log-normal function, which means the event rate of solar flare changes randomly. 

Regarding the waiting time of FRBs, the earlier assumption was a non-stationary Poisson process or a Weibull distribution \cite{2017JCAP...03..023W, 2018MNRAS.475.5109O}, albeit influenced by the sparse sampling of early burst events. As depicted in Fig. \ref{fig:wt}, the bimodal waiting time distribution of FRBs evidently differs from a single Weibull distribution, which can be well applied to earthquake events. There are also diverse models used to describe the distribution of waiting times for FRBs, including LogNormal distribution \cite{2022MNRAS.515.3577H}, Weibull \cite{2021MNRAS.500..448C}, time-dependent Poisson processes \cite{2023MNRAS.519..666J, 2023ApJ...949L..33W}.

For comparative analysis, using FRB 20121102A as an example, we selected waiting times exceeding 1 second. Fitting was undertaken with exponential (corresponding to a Poisson process), Weibull, and lognormal functions. The goodness of fit was assessed using adjusted R-squared, which provides a refined assessment of the model's fitting effectiveness by accounting for the number of independent variables utilized. The parameter counts for exponential, Weibull, and lognormal are 1, 2, and 2, respectively. The calculated adjusted R-squared values were 0.920, 0.917, and 0.906 for these three functions. While all three functions appear adept at modeling FRB waiting times, we lean toward the simplest model, the exponential (Poisson process) function, for describing the distribution of FRB waiting times. The time-dependent Poisson process, which is the sum of multiple Poisson processes, entails a greater number of parameters and evidently excels in fitting residuals compared to models with fewer parameters. Nevertheless, under the simplest conditions, our results indicate the effectiveness of a single Poisson process. Consequently, in this context, we opt for employing two Poisson processes to fit the entire distribution of FRB waiting times.

We also explored the distribution of waiting times under different energy thresholds. Taking FRB 20121102A as an example, we set a series of energy thresholds, calculated the waiting time distribution for each, and employed an exponential function to fit the two peaks of waiting times. For exponential distributed waiting times, $P(t)=\lambda e^{-\lambda t}\, {\rm d} t=\lambda e^{-\lambda t} t\, {\rm d} \log t$, the peak corresponds to the position where the derivative is zero $P'(t)=0\Rightarrow t=1/\lambda$. Thus, the exponential distribution provides the typical waiting time corresponding to the peak. In Fig. \ref{fig:wt}g-h, we present the characteristic times corresponding to the two peaks of waiting time under different energy thresholds. The left peak (short timescale waiting times) is insensitive to the choice of energy threshold, implying an intrinsic correlation with the radiation mechanism. The right peak (long timescale waiting times) strongly depends on the energy threshold or the detected event number. As the energy threshold increases, it leads to a decrease in the detected bursts, resulting in longer waiting times. In panel H, two inflection points correspond to the two peaks in the energy distribution. After normalizing the right peak's waiting time using the detected number of bursts under different energy thresholds and observation duration, we can see that all the normalized right peak hovers around 1, indicating that the right peak is a stochastic process relying on sampling (event rate). When the waiting time of FRBs can be described by stochastic processes, it further disfavors the prospect of (quasi)-periodicity in FRBs. 

The distinct characteristics of the waiting time distribution underscore that the emission of FRBs is not like seismic events. A recent study has proposed a semblance of aftershock characteristics between FRBs and earthquakes through a custom correlation function \cite{2023MNRAS.526.2795T}. However, the similarity in the ``correlation function'' between earthquakes and FRBs arises from the bias of waiting time distributions from one Poisson process. Here, we show that despite both earthquakes and FRBs deviating from a Poisson process in terms of waiting times, the manner of deviation differs. Therefore, it is hard to conclude that FRBs exhibit seismic aftershock characteristics. This assertion is further substantiated in our subsequent analysis of energy distribution.

\section{Energy domain analysis and results}

\begin{figure*}[!htp]
    \centering
    \includegraphics[width=0.9\textwidth]{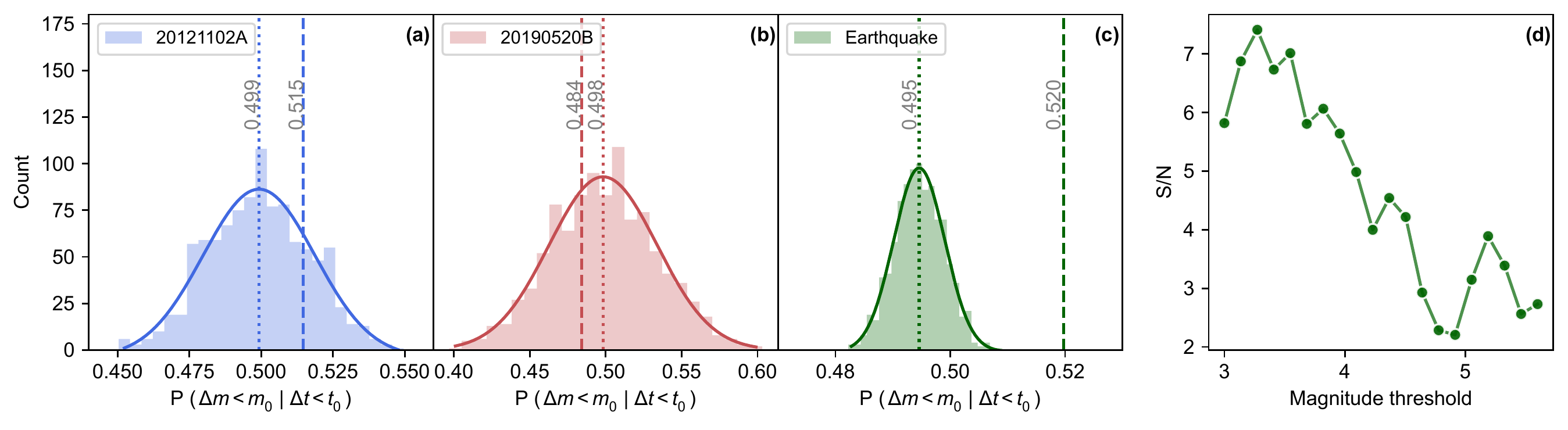}
    \caption{{\bf The probability (P) of the hypothesis ``the energy of the subsequent event is lower than that of the preceding event''.} {\bf Panel a-c} show this probability (P) of the original event sequences (dashed lines) for FRB 20121102A (blue), FRB 20190520B (red), and earthquakes (green). The histogram and the dotted lines show the distribution and median of this probability (P) of randomly shuffled event sequences. {\bf Panel d} illustrates the variation in the S/N of the probability (P) under different earthquake magnitude thresholds.}
    \label{fig:pmt}
\end{figure*}

Despite the common perception of randomly occurring earthquakes, seismic events cluster in time and magnitude \cite{omori1895after, 1980GeoJ...62..303K, eqcluster}. The deviation of earthquakes from randomness origins from its nonlinear dynamical systems \cite{omori1895after, 2007JGRB..112.4313S}. Assuming in an event series, the subsequent event often possesses lower energy than its predecessor. After shuffling, the probability of subsequent energy being lower than the predecessor will decrease. Based on this fact, we adopted the analysis method used in \cite{eqcluster} for earthquakes.

Consider conditional probability
\begin{equation}
    P(\Delta m_i < m_0 | \Delta t_i < t_0) = \frac{N(\Delta m_i < m_0, \Delta t_i < t_0)}{N(\Delta t_i < t_0)} \>,
\end{equation}
where $\Delta t_i=t_{i+1}-t_i$ and $\Delta m_i=m_{i+1}-m_i$ are the temporal and magnitude difference between subsequent events, $N(\cdots)$ is the number of pairs of subsequent events satisfying the conditions specified in parentheses. 

With fixed $m_0$ and $t_0$, the quantity $P(\Delta m^*_i < m_0 | \Delta t_i < t_0)$ for independent reshuffled series are calculated, where $\Delta m^*_i=m^*_{i+1}-m^*_{i}$ is the magnitude difference between subsequent events of reshuffled series. We employed \textsc{numpy.random .shuffle} to rearrange our event series, which is based on the Fisher-Yates shuffle algorithm to produce an unbiased permutation. The central limit theorem makes the quantity $P(\Delta m^*_i < m_0 | \Delta t_i < t_0)$ a Gaussian distribution with mean value $\mu(m_0,t_0)$ and standard deviation $\sigma(m_0,t_0)$. The biases between $\mu(m_0,t_0)$ and $P(\Delta m_i < m_0 | \Delta t_i < t_0)$ for earthquake, FRB~20121102A and FRB~20190520B are $5.65\sigma$, $0.79\sigma$ and $0.40\sigma$ (Fig.~\ref{fig:pmt}a-c), respectively. Generally, the detection of earthquakes is more complete than that of FRBs. Here, we conducted a test related to earthquake magnitude thresholds. For FRB 20121102A, the burst with the highest energy differs by a factor of 4000 from the one with the lowest energy, and for FRB 20190520B, the factor is 500. In the earthquake data, the highest magnitude is 7.5. By progressively increasing the energy threshold, we made the lowest-energy earthquake event reach 1/500 of the highest-energy event. Throughout this process, we calculated the S/N of the hypothesis ``the energy of the subsequent event is lower than that of the preceding event'' for different thresholds. In Fig. \ref{fig:pmt}d, we can see that even when compressing the dynamic range of earthquake energies to 1/500, the S/N still exceeds 2, far surpassing the two FRBs. This indicates that no clustering in the energy of FRB bursts. Simply put, there is no excess number of bursts following a bright one, while there are more after-quakes following a major event than pre-quakes. 

\section{Stochasticity and Chaos}

In dynamical studies, chaos and stochasticity are two distinct concepts. Chaos is characterized by unpredictability that increases with time, whereas stochasticity's unpredictability remains stable over time. We use ``Pincus Index'' (\pin) and ``Maximum Lyapunov exponent'' (\mle), respectively, to quantify the stochasticity and chaos of event sequences.

The \pin is used to describe the degree of stochasticity based on the Max Approximate Entropy (\mae) \cite{1991PNAS...88.2297P, entropy}, by measuring the change in information entropy before and after shuffling a sequence. \pin is zero for completely ordered systems and one for totally random systems. For an event series $\{e_i\}_{i=1,\cdots, N}$, the \mae \cite{1991PNAS...88.2297P, entropy} can be defined as

\begin{equation}
    {\mathcal{MAE}} = \max_{r}\left(\left.-\frac{1}{N-m}\sum_{i=1}^{N-m}\log \frac{\sum_{j=1}^{N-m}{\rm dist}(x_j, x_i)<r}{N-m}\right|^{m+1}_m\right) \>,
\end{equation}
where $X = \{x_{i}\}_{i=1,\cdots,N-m} = \{\{e_1,\cdots, e_{1+m}\}, \cdots, \{e_{N-m}, \cdots, \\ e_{N}\}\}$ is the reorganized sequence of the initial event series, $N$ is the length of the initial event series, ${\rm dist}(x_j, x_i)$ is the distance between each pair of the reorganized sequence $X$, and $r$ is the distance threshold. \mae quantifies the greatest difference in information between the segments of length $m$ and $m+1$, which needs to obtain a max entropy of the recombination sequence by changing different thresholds $r$. For making the \mae comparable between series, the \pin was defined as
\begin{equation}
    {\mathcal PI} = \frac{\mathcal{MAE}_{\rm initial}}{\mathcal{MAE}_{\rm shuffled}} \>.
\end{equation}
After calculating the \mae or called ${\mathcal{MAE}_{\rm initial}}$ for initial event series, we randomly reshuffled the series and calculated \mae for 100 times. The median value is ${\mathcal{MAE}_{\rm shuffled}}$, the standard deviation value is used to define the error of \pin. In the calculation of the \pin, the distance metric used is the Euclidean distance. To preserve the original sequence information as much as possible and minimize the introduction of artificial bias, we linearly mapped both the time and energy sequences to the range of 0 to 1 simply.

The \mle represents the degree of dispersion of trajectories in phase space and is a numerical characteristic used to identify chaotic behavior in a nonlinear system \cite{lyp}. \mle less than 0 corresponds to a periodic motion or a stable system that is static in time-energy space. \mle greater than 0 indicates the existence of chaos. We use \textsc{nolds} \cite{nolds} method for \mle calculation. As the \mle is the maximum value of the whole spectrum of Lyapunov exponents, it is difficult to define an error.

\begin{figure*}[!htp]
    \centering
    \includegraphics[width=0.7\textwidth]{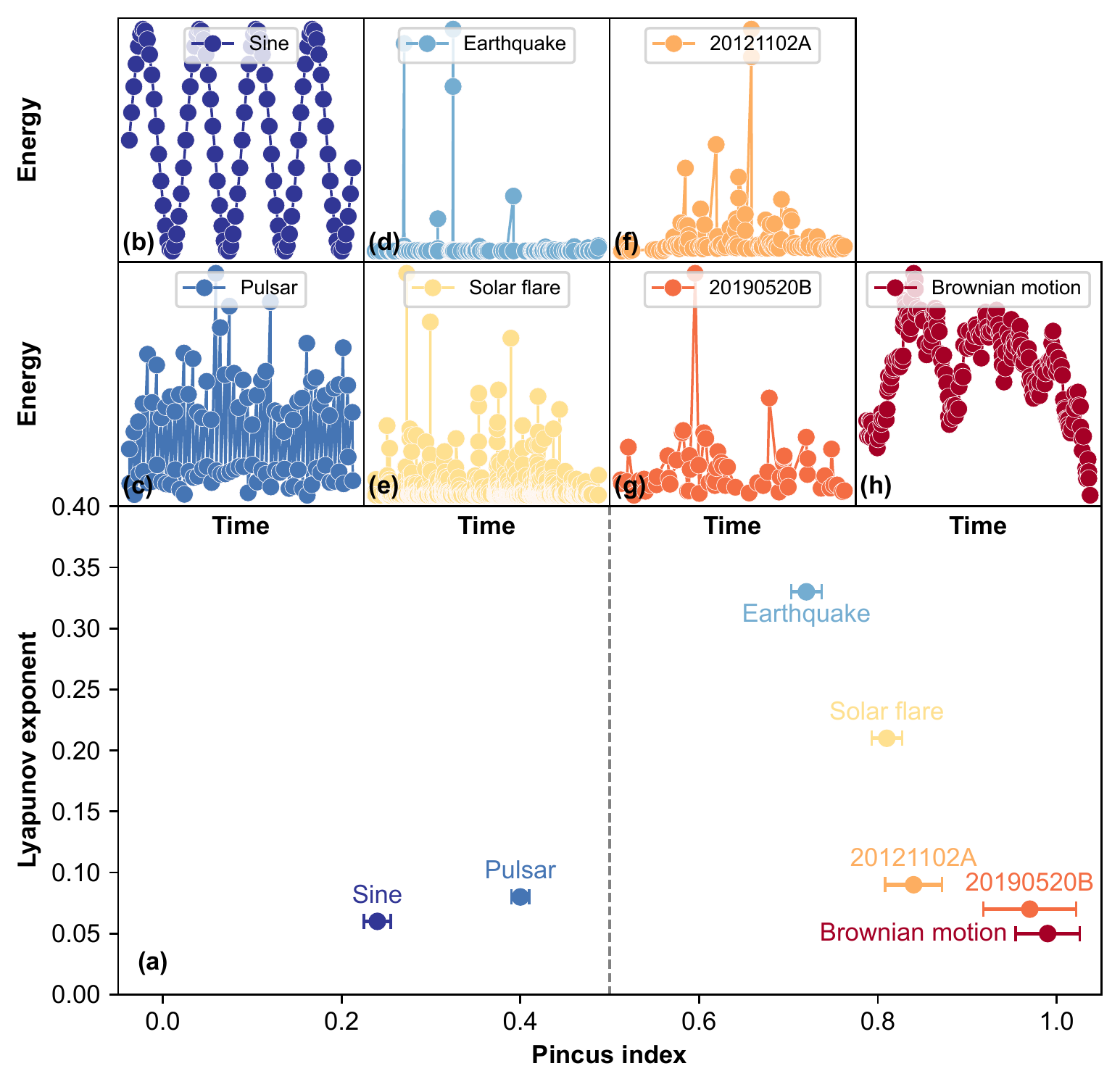}
    \caption{{\bf The Pincus index vs. Lyapunov exponent of sine function, pulsar, earthquake, solar flare, FRB~20121102A, FRB~20190520B and Brownian motion.} The top and middle panels present event series in time-energy space of these sources. The color changes from blue to red, implying increased stochasticity. (\pin, \mle) value is (0.24, 0.06) for sine-wave, (0.40, 0.08) for pulsar, (0.72, 0.33) for Earthquake, (0.81, 0.21) for solar flare, (0.84, 0.09) for FRB~20121102A, (0.97, 0.07) for FRB~20190520B, and (0.99, 0.05) for Brownian Motion.}
    \label{fig:pi}
\end{figure*}

The \pin and \mle values are shown in Fig.~\ref{fig:pi}a. The pulsar data used here was a pulsar named J1840+2843 discovered within the CRAFTS project \footnote{\url{http://groups.bao.ac.cn/ism/CRAFTS/202203/t20220310_683697.html}}. 
As expected, the most regular motion is a sine-wave, with small values in both, followed by pulsar with representative \pin$\sim$0.4 and \mle$\sim$0.08.
Earthquakes are most chaotic with \mle$\sim$0.33. 
Solar flares are less chaotic, but more random than earthquakes. 
FRBs are even more random (\pin$\sim$0.84-0.97) and less chaotic (\mle$\sim$0.07-0.09) than solar flares, mimicking Brownian Motion (\pin$\sim$0.99) that is the most random in the bivariate time-energy space, among the systems considered here. 

\begin{figure}[!htp]
    \centering
    \includegraphics[width=0.47\textwidth]{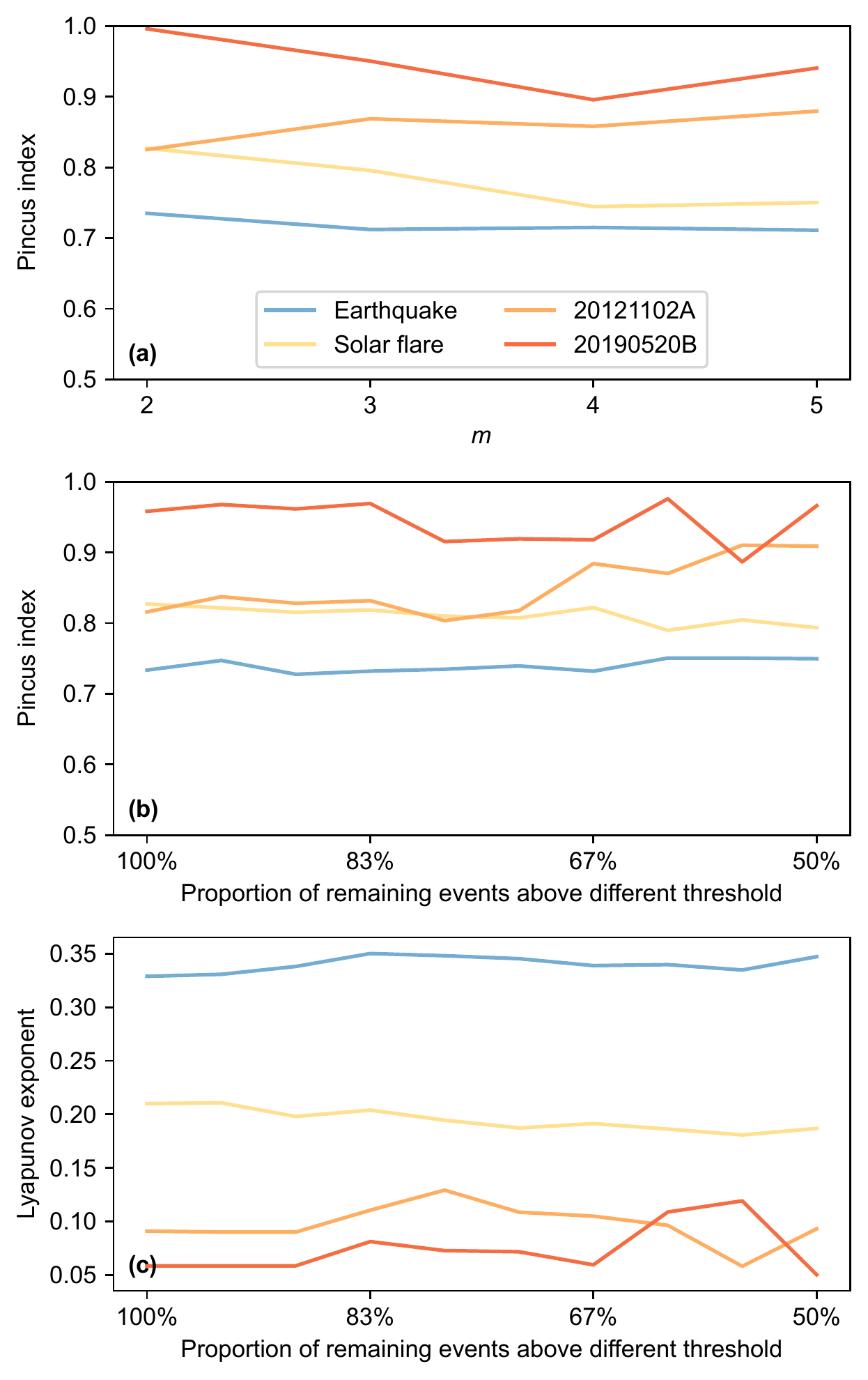}
    \caption{{\bf The robustness test of \pin and \mle.} {\bf Panel a} shows the \pin values using different segments of length $m$. {\bf Panel b-c} show the \pin and \mle under different energy thresholds.}
    \label{fig:test}
\end{figure}

To ascertain the robustness of our calculations, we considered two types of tests. In the calculation of the \pin, only the parameter $m$ remains as a free variable. We conducted tests by adopting $m \in \{2, 3, 4, 5\}$. As depicted in Fig.~\ref{fig:test}a, the deviation of the \pin did not exceed 0.1 for different values of $m$. Even with slight variations in the \pin, the relative relationships among different physical phenomena remained unchanged when the same $m$ was selected. Additionally, given that the detection of these physical phenomena cannot be exhaustive, we further examined the impact of sequence completeness on the results of the \pin and \mle calculations. By adjusting the energy threshold, we retained different proportions of events (ranging from 100\% down to 50\%) and subsequently computed the \pin and \mle for each case. From retaining 100\% of the events to selecting only those with the top 50\% energy levels, the \pin and \mle exhibited remarkable stability (Fig. \ref{fig:test}b-c), confirming the robustness of our calculations.

Within the stochasticity-chaos phase space, FRBs are noticeably distinct from earthquakes, exhibiting lower levels of chaos and higher levels of stochasticity. Additionally, as previously analyzed, earthquakes exhibit time and energy clustering behavior (Figs.~\ref{fig:wt}-\ref{fig:pmt}). FRBs, however, do not show clustering in neither time nor energy. The random paths of FRB events in time-energy space favor origin models based on complex processes, such as in a thermodynamic system, the movement or diffusion of molecules is akin to the phenomenon of interest. 

\section{Conclusion}

In summary, using \pin and \mle to quantify the stochasticity and chaos of event sequences may be an effective method for intuitively demonstrating the similarities and differences between various physical processes. Specifically, here we compare FRBs, pulsar, solar flares, earthquakes, and Brownian motion in the stochasticity-chaos phase space. The strong stochasticity, akin to Brownian motions, of active repeaters with PRS counterparts, along with the growing evidence of their multi-variate behaviors, such as their bimodal energy distribution\cite{li21, 2021ApJ...919...89Y, 2022RAA....22l4002Z, 2023ApJ...955..142Z}, could be generated by a single source with high information entropy or the combination of multiple radiation mechanisms or emission sites. Either way, it is unlikely that active repeating FRBs originate from a stably spin compact objects within a clean environment.

\section*{Competing interests}
The authors declare no competing interests.

\section*{Acknowledgments}
This work made use of the data from FAST (Five-hundred-meter Aperture Spherical radio Telescope). FAST is a Chinese national mega-science facility, operated by National Astronomical Observatories, Chinese Academy of Sciences. This work supported by the Open Project Program of the Key Laboratory of FAST, Chinese Academy of Sciences. The Parkes radio telescope (Murriyang) is part of the Australia Telescope National Facility, which is funded by the Commonwealth of Australia for operation as a National Facility managed by CSIRO. This work was partly carried out by using Hinode Flare Catalogue (\url{https://hinode.isee.nagoya-u.ac.jp/flare_catalogue/}), which is maintained by ISAS/JAXA and Institute for Space-Earth Environmental Research (ISEE), Nagoya University. Yong-Kun Zhang thanks Wei-Yang Wang for the comments about the star-quake model. 

Di Li and Yi Feng are supported by NSFC grant No. 11988101, 11725313, 11690024, 12203045, and by Key Research Project of Zhejiang Lab No. 2021PE0AC03. Di Li is a New Cornerstone Investigator. Pei Wang is supported by NSFC grant No. U2031117, the Youth Innovation Promotion Association CAS (id.~2021055) and the Cultivation Project for FAST Scientific Payoff and Research Achievement of CAMS-CAS. Chen-Hui Niu is supported by NSFC grant No. 12203069, the National SKA Program of China/2022SKA0130100 and the Office Leading Group for Cyberspace Affairs, CAS (No. CAS-WX2023PY-0102). Shi Dai is the recipient of an Australian Research Council Discovery Early Career Award (DE210101738) funded by the Australian Government. Ju-Mei Yao is supported by the National Science Foundation of Xinjiang Uygur Autonomous Region (No. 2022D01D85), the Major Science and Technology Program of Xinjiang Uygur Autonomous Region (No. 2022A03013-2), the Tianchi Talent project and the CAS Project for Young Scientists in Basic Research (grant YSBR-063).

\section*{Author contributions}
Yong-Kun Zhang and Di Li developed the concept of the manuscript. Yong-Kun Zhang conducted the data analysis and visualization. Pei Wang, Chen-Hui Niu, Shi Dai, and Yong-Kun Zhang searched the new bursts and analysed the burst properties. Yong-Kun Zhang, Di Li, Chao-Wei Tsai, Ju-Mei Yao, and Yi Feng led the discussion on the interpretation of the results and writing of the manuscript. All authors contributed to the analysis or interpretation of the data and to the final version of the manuscript.
\printcredits

\section*{Data availability}
Earthquake data from the Southern California Earthquake Data Center is used in this paper. The full dataset and documentation can be downloaded from \url{https://dx.doi.org/10.7909/C3WD3xH1}. Solar flare data from Hinode Flare Catalogue is used in this paper. The full dataset and documentation can be downloaded from \url{https://doi.org/10.34515/CATALOG.HINODE-00000}. Other data can be accessed in ScienceDB \url{https://doi.org/10.57760/sciencedb.09716}.

\bibliographystyle{scibull}
\bibliography{cas-refs}

\end{document}